\documentclass[a4paper]{jpconf}
\usepackage{citesort}

\usepackage{graphicx}
\usepackage{bm}

\def\d#1{\,{\rm d}#1}
\def\e#1{\,{\rm e}#1}

\def\vec#1{{\overrightarrow{#1}}}

\def\and{\,\,\&\,}

\def\ket#1{|#1\rangle }
\def\bra#1{\langle #1|}

\def\braket#1#2{\langle #1|#2\rangle}

\def\i{{\bf i}}

\def\skipm#1{}

\def\vec#1{{\bm{#1}}}
\def\figurehight{3.2in}
\def\paragraph#1{}
\begin{document}
\title{Non-adiabatic effect on Laughlin's argument 
of the quantum Hall effect
}
\author{I. Maruyama$^1$ and Y. Hatsugai$^2$}
\address{
$^1$Department of Applied Physics, University of Tokyo, Hongo Bunkyo-ku, Tokyo 113-8656, Japan\\
$^{2}$Institute of Physics, Univ. of Tsukuba, 1-1-1 Tennodai, Tsukuba Ibaraki 305-8571, Japan
}
\ead{maru@pothos.t.u-tokyo.ac.jp}

\begin{abstract} 
We have numerically studied a non-adiabatic charge transport
in the quantum Hall system pumped by a magnetic flux,
as one of the simplest theoretical realizations of non-adiabatic Thouless pumping.
In the adiabatic limit,
a pumped charge is quantized,
known as Laughlin's argument in a cylindrical lattice.
In a uniform electric field, we obtained a formula connecting quantized pumping in the
adiabatic limit and no-pumping in the sudden limit.  
The intermediate region between the two limits is determined by  the Landau
gap.
A randomness or impurity effect is also discussed.
\end{abstract}

\paragraph{Laughlin}
In the
paper by Laughlin\cite{PRB.23.5632}, the quantum Hall system on a cylinder with two edges
penetrated by an Aharonov-Bohm (AB) flux $\Phi$ is considered, where
the flux changes from 0 to a flux quantum $\Phi_0=h/e$ {\it adiabatically}.  
The change of the AB
flux enforces electrons to move form one edge to the other edge.  
In
the adiabatic limit, net charge of transported electron is quantized
due to a request from the gauge transformation, which is known as Laughlin's
argument in a cylindrical lattice.  
This topological aspect is the key feature of quantization of the Hall conductivity\cite{PRB.25.2185,PRL.49.405,PRL.71.3697}.

\paragraph{Thouless Pumping}
Although Laughlin's argument gives a theoretical explanation for
the quantum Hall effect, quantized charge transport is realized in various situations as
Thouless pumping\cite{PRB.27.6083}.
Thouless argued that an electron system in time dependent potential
such as right moving potential $U(x,t)=\sin\left(2\pi (x/L-t/T)\right)$  with period $T$ and $L$
can pump a quantized charge
in analogy with water pumping by Archimedean screw\cite{SCIENCE.283.1864}.
Recently, in mesoscopic system,
electron pumping by adiabatic change of a cyclic potential $U (x,t)$ has
attracted much attention both experimentally\cite{SCIENCE.283.1905}
and theoretically\cite{SCIENCE.283.1864}.
Not only adiabatic pumping
but also
non-adiabatic pumping is also important and
realized easily in experimental situation\cite{PRB.77.153301},
which needs a extension of original Thouless pumping
theoretically\cite{PRB.66.205320}.
To increase net current induced by successive pumping,
non-adiabatic pumping is thought to be more efficient 
because 
fast pumping transports more electrons\cite{PRL.95.130601}.
\paragraph{in this...}
\begin{figure}
\begin{minipage}{8cm}
\resizebox{6cm}{!}{\includegraphics{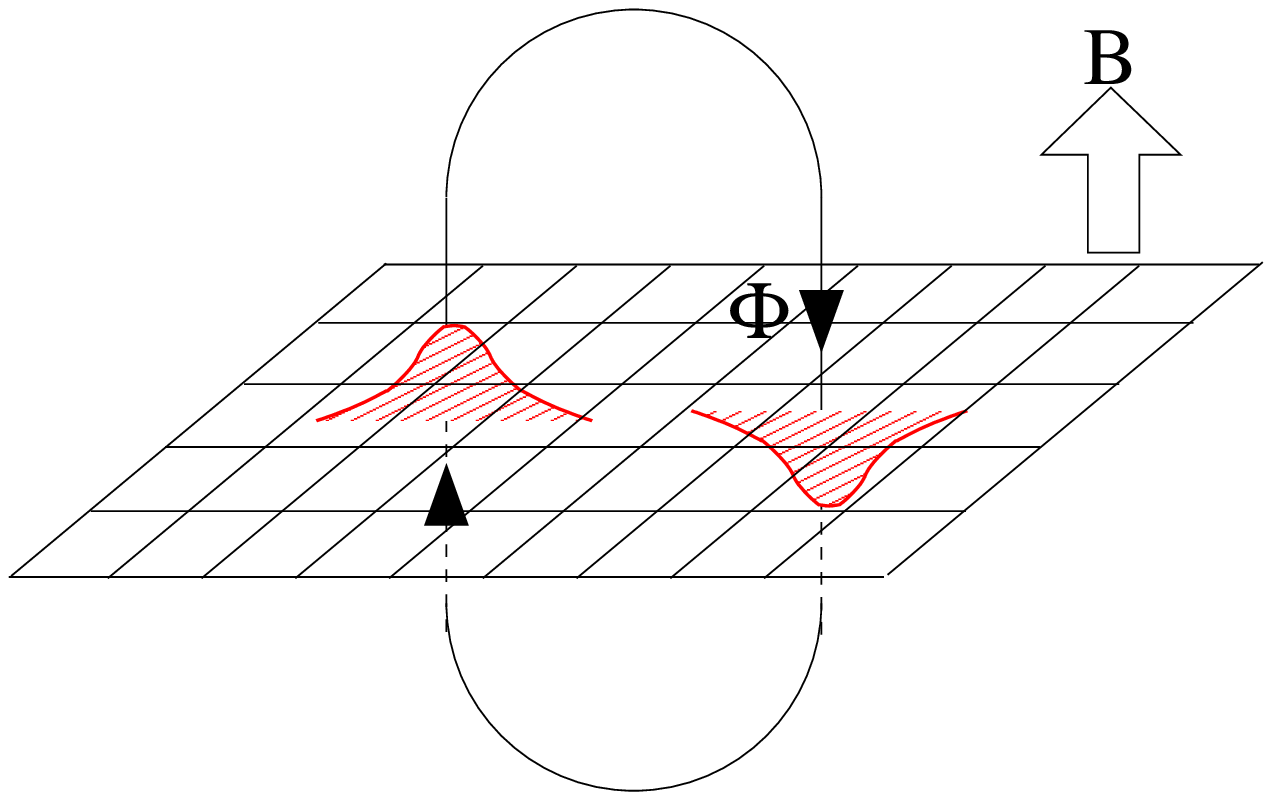}}
\caption{Schematic figure of edge-state  induced by a magnetic flux $\Phi$
in the quantum Hall system.
}
\label{fig:Ab}
\end{minipage}\hspace{2pc}%
\begin{minipage}{8cm}
\resizebox{6cm}{!}{\includegraphics{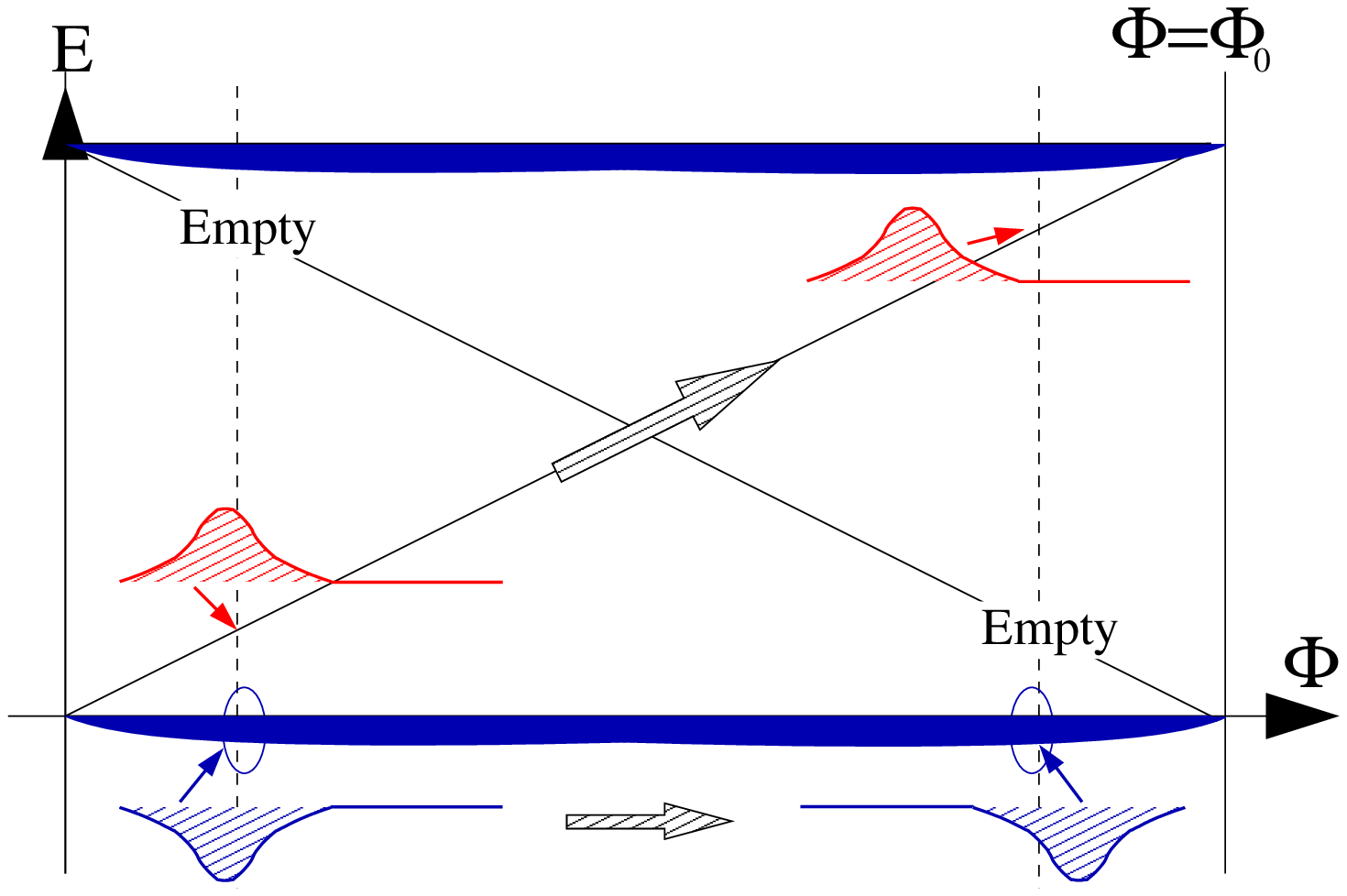}}
\caption{Energy diagram of edge-state pumping 
as a function of a magnetic flux $\Phi$.
Each inset with a shaded region indicates local density.
}
\label{fig:Ac}
\end{minipage}
\end{figure}
In this paper,
going back to the cylindrical system of Laughlin's argument,
we change the AB flux $\Phi$ {\it non-adiabatically},
i.e.,
non-adiabatic effect on edge-state pumping.
For this purpose,
we introduce a time-dependent flux $\Phi(t)=\Phi_0 t/T$ with the period $T$.
This system is one of the simplest theoretical realizations of non-adiabatic Thouless pumping.
In addition,
we shall study a square lattice penetrated by the flux $\Phi$ without boundary
as shown in Fig.~\ref{fig:Ab}.
There is no edge state at $\Phi=0$ and edge states induced by $\Phi$.
One of our motivations is
to
study how a edge state goes through $\Phi = \Phi_0$
because
the edge state
comes across the second Landau level
as shown in Fig.~\ref{fig:Ac}.
Here we suppose the lowest Landau level is filled at $\Phi=0$.
Level crossing of left and right edge states at $\Phi = \Phi_0/2$ 
is easily understood by a Landau-Zener tunneling as studied massively in one-dimensional ring\cite{PRB.33.6497}.

\paragraph{method}
We investigate the time evolution of the ground state 
in a square lattice with the lengths $L_x$ and $L_y$ 
under the magnetic field $B a^2= p/q \Phi_0$,
where $a$ is the lattice constant.
The time-dependent Schr\"odinger equation $\i \partial_t \ket{\Psi(t)} = H \ket{\Psi(t)}$
is solved numerically with time step $\Delta t$.
To preserve the norm $\braket{\Psi(t)}{\Psi(t)}$ numerically,
the Suzuki Trotter decomposition of $H$ is used\cite{EPL.3.139,PRB.51.10897}.
Although $H$ is one-body Hamiltonian,
the state $\ket{\Psi(t)}$ is a many-body state
filled up to Fermi energy which is fixed around the center of the first Landau gap.
To observe pumped charge,
we calculate $\Delta N = \bra{\Psi(t)} (N_R - N_L) \ket{\Psi(t)}$,
where $N_L$ ($N_R$) is the number of electrons of left (right) system.
$\Delta N(T)=0$ in the sudden limit
while $\Delta N(T)=-2$ in the adiabatic limit
because the left edge state is occupied
and the right edge state is empty at $t=T$ as shown in Fig.~\ref{fig:Ac}.
In other words,
charge is carried by extended states of the first Landau band as explained in
Laughlin's argument.
Here after, we limit ourselves to the commensurate lattice with
the periodic boundary condition (PBC):
$L_y = q\times l_y$, $L_x = q\times l_x$
with integer numbers $l_x$ and $l_y$.
We set 
the magnetic field $p/q=1/7$, 
the unit of energy $t=1$,
$a=1$,
and
the system size and time step are took large enough
to obtain the thermodynamic limit.
The maximum lattice size is $L_y=L_x=70$.

\begin{figure}[h]
\begin{minipage}{3in}
\resizebox{3in}{\figurehight}{\includegraphics{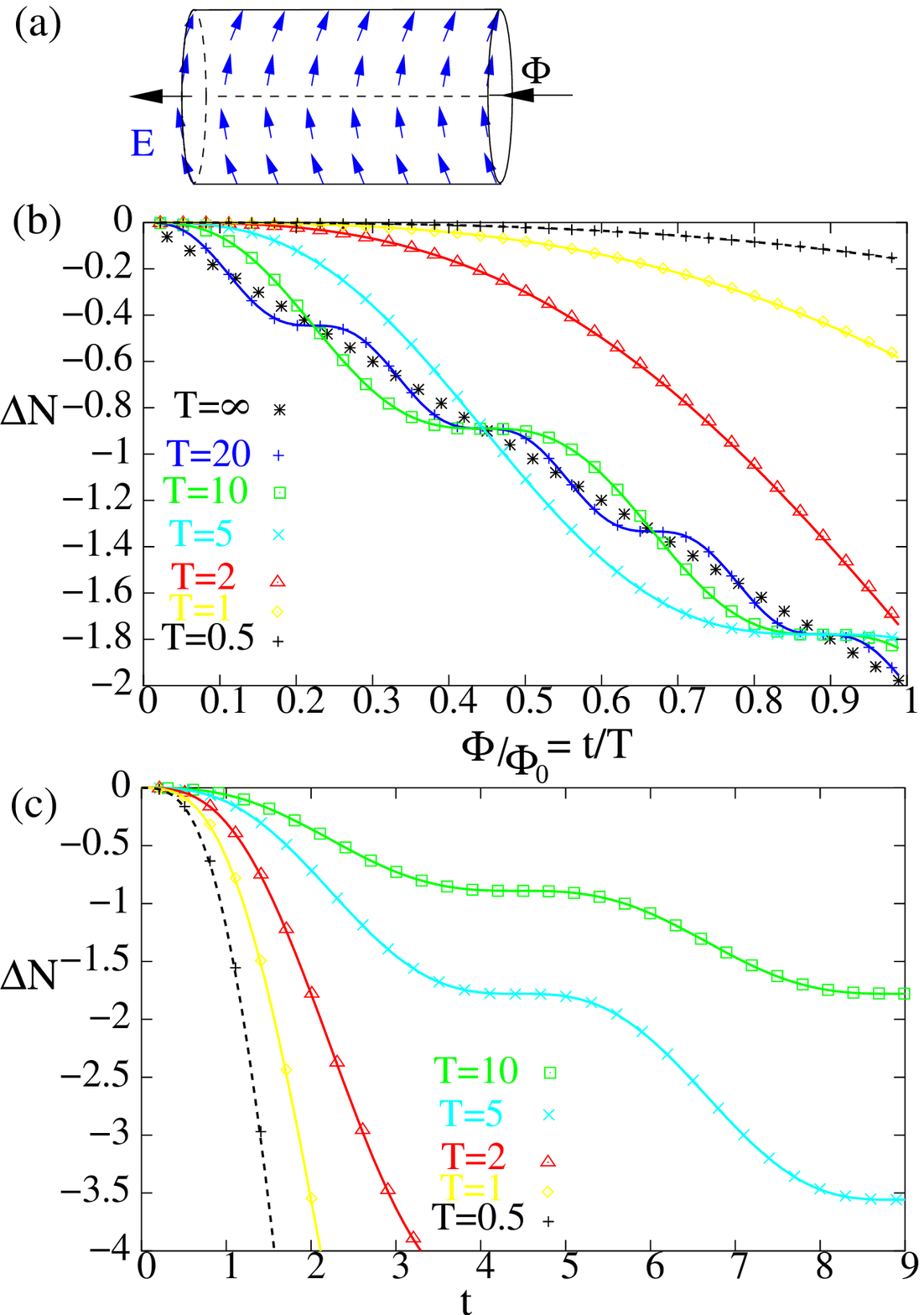}}
\caption{(a) model of $H_{uni}$ on the cylindrical lattice
(b) $\Delta N$ as a function of $\Phi/\Phi_0$.
Solid lines are $\Delta N$ of Eq.~\ref{eq:DeltaN}
with fixed $\omega_c$
and respective $T$.
(c) $\Delta N$ as a function of $t$.

}
\label{fig:B}
\end{minipage}\hspace{2pc}%
\begin{minipage}{3in}
\resizebox{3in}{\figurehight}{\includegraphics{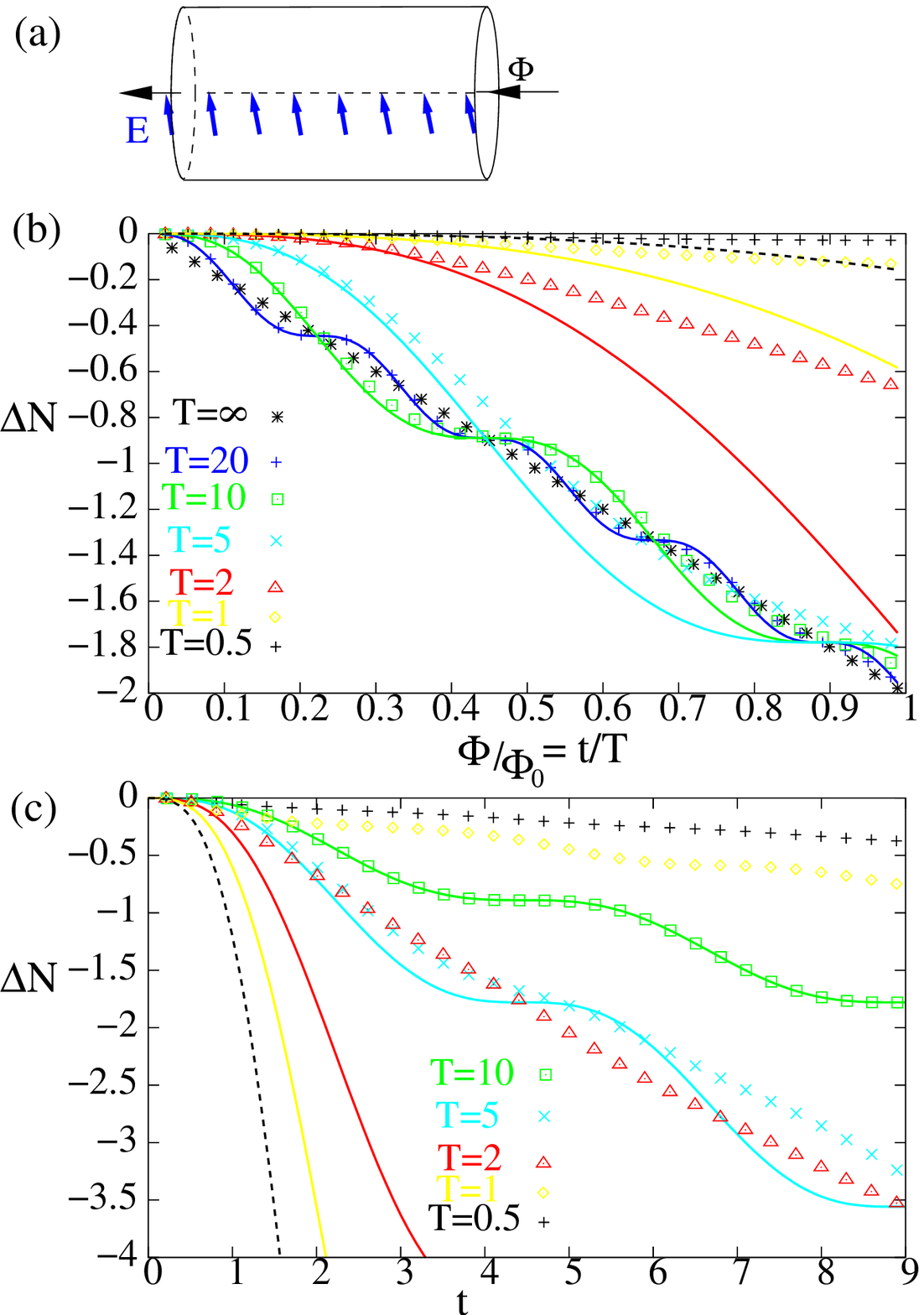}}
\caption{(a) model of $H_{imp}$ on the cylindrical lattice
(b) $\Delta N$ as a function of $\Phi/\Phi_0$
Solid lines are $\Delta N$ of Eq.~\ref{eq:DeltaN}
with fixed $\omega_c$
and respective $T$.
(c) $\Delta N$ as a function of $t$
}
\label{fig:C}
\end{minipage}
\end{figure}

\paragraph{model}
The Hamiltonian is defined as
\begin{math}
  H(\Phi) = -t\sum_{mn}   c_{m+1,n}^\dagger c_{mn} 
  -t \sum_{mn} c_{m,n+1}^\dagger \e^{\i \theta_{m n}^y} c_{mn}
  + h.c.
,
\end{math}
where we take the Landau gauge
$\theta_{m n}^x =0$.
When we consider the cylindrical system,
a site index $(m,n)$ is limited to $m\in[1,L_x-1], n\in[1,L_y]$.
For $\theta_{m n}^y$, we consider two Hamiltonians;
$H_{uni}(t)$ under uniform electric field and
$H_{imp}(t)$ under local electric field,
where the electric field is given by $\vec{E}:=-\partial_t \vec{A}$
and vector potential is related with the hopping phase through
$\theta_{mn}^y=\int_{\vec{r}_{mn}}^{\vec{r}_{m,n+1}} \vec{A}\cdot\d{\vec{r}}$.
Note that the dynamic electric field can be determined by the electron system
but we suppose the static electric field
to consider the electron system in two simple limits.
The uniform electric field shown in Fig.~\ref{fig:B}(a)
is realized by 
\begin{math}
  \theta_{m n}^y = 2\pi (p m/q + {t\over T L_y})   
\end{math}
and the local electric field shown in Fig.~\ref{fig:C}(a)
is realized by 
\begin{math}
  \theta_{m n}^y =2\pi (p m/q + \delta_{nL_y} {t\over T})   
.
\end{math}
Except in the adiabatic limit,
there is no static gauge transformation between $H_{uni}$ and $H_{imp}$.
Of course, one can find the function $\chi(t,\vec{r})$ of the time-dependent gauge transformation $\vec{A}_{imp}(t,\vec{r}) = \vec{A}_{uni}(t,\vec{r})-\vec{\nabla} \chi(t,\vec{r})$
with taking into account new scalar potential $\phi(t,\vec{r})=-\partial_t \chi(t,\vec{r})$.
That is, $H_{uni}$ and $H_{imp} + \phi(t,\vec{r})$ are the same.
Although $H_{uni}$ is more natural than $H_{imp}$,
we shall also study $H_{imp}$ for comparison.
Note that $H_{uni}(t+T)\neq H_{uni}$ and $H_{imp}(t+T)=H_{imp}(t)$.

\paragraph{Uniform case}
When we consider the cylindrical lattice described by $H_{uni}$,
a wave number in the $y$ direction is a good quantum number
and preserved for any $T$.
Then, the possible diabatic transition occurs in each separated sector labeled by $k_y$.
After the Fourier transformation
\begin{math}
  c_{m,n} = {1\over L_y} \sum_{k_y} \e^{\i k_y n} c_m(k_y)
  ,
\end{math}
we get
\begin{math}
  H_{uni}= {1\over L_y} \sum_{k_y} H(k_y)
\end{math}
with 
\begin{math}
  H(k_y) = -t \sum_{m=1}^{L_x-2} \left(c^\dagger_{m+1}(k_y) c_m(k_y) + h.c.\right)
  -2t\sum_{m=1}^{L_x-1} U(m,t) c^\dagger_m(k_y)c_m(k_y)
,
\end{math}
where the on-site potential $U(m,t)=\cos\left(k_y - 2\pi (p m/q   + {t\over T L_y})\right)$
is periodic: $U(m+q,t)=U(m,t+T L_y)=U(m,t)$.
This Hamiltonian is one of the simplest theoretical realizations of Thouless pumping.

\paragraph{Result}
As a result, the spatial distribution of the charge density shows the
quantized charge transport, i.e, $\Delta N(T)=-2$
in some parameter range.
There arises the question 
when quantization of pumped charge breaks down,
i.e.,
the limitation of Laughlin's argument in a non-adiabatic process.
Considering the Landau-Zener tunneling,
the upper limit of $T$ 
is given as $T<\hbar/ \Delta_e$,
where $\Delta_e$ is an energy gap of the overlapped edge states due to finite $L_x$.
Since $\Delta_e$ is exponentially small due to
large system size,
the upper limit of $T$ becomes infinite.
On the other hand,
the lower limit of $T$ is $1 / \omega_c$,
where $\hbar \omega_c$ is the Landau gap.
Figure ~\ref{fig:B}(b)
shows $\Delta N$ as a function of $\Phi/\Phi_0 = t/T$.
At large $T>1$
pumped charge is quantized $\Delta N(T)=-2$,
while $\Delta N(T)$ becomes zero at small $T$.
Solid lines are fitted with the formula
\begin{eqnarray}
\Delta N = -{2t\over T} + {2\over \omega_c T} \sin(\omega_c t)  
\label{eq:DeltaN}
,
\end{eqnarray}
which connects quantized pumping in the
adiabatic limit $T=\infty$ and no-pumping in the sudden limit $T=0$
and 
actually shows good agreement as shown in Fig.~\ref{fig:B}(b) and (c).
We note that $\omega_c$ is about 1.4.
However,
Fig.~\ref{fig:C} for $H_{imp}$ 
shows disagreement with solid lines given by Eq.~\ref{eq:DeltaN}
especially for small $T$.
A major difference
is shown in Fig.~\ref{fig:B}(c) and Fig.~\ref{fig:C}(c).
Pumped charge per time can increase by fast pumping only in the uniform system.

\paragraph{Effect of randomness}
\begin{figure}[h]
\begin{minipage}{3in}
\resizebox{3in}{!}{\includegraphics{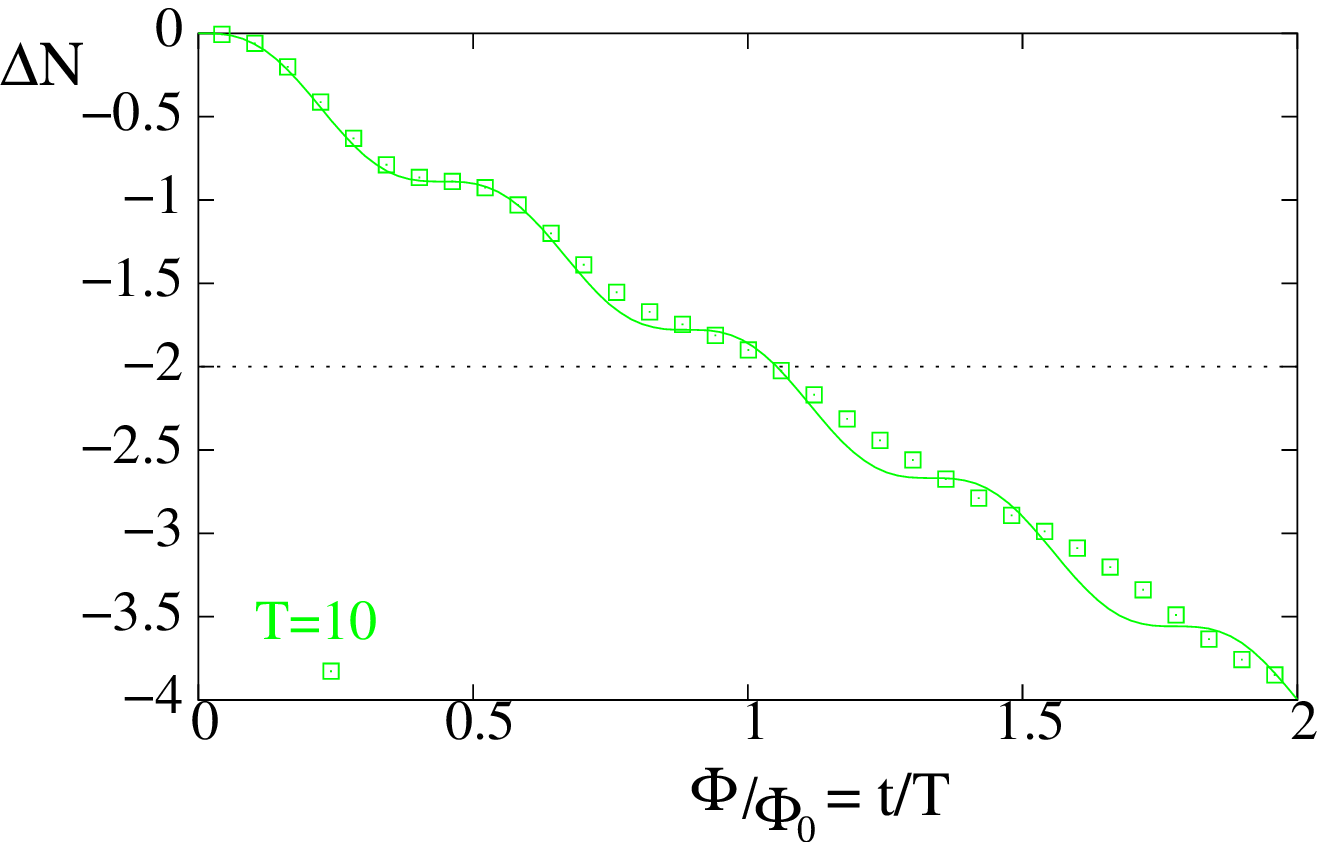}}
\caption{$\Delta N$ as a function of $\Phi/\Phi_0$
for $H_{uni}$ with the PBC at $T=10$.
}
\label{fig:D}
\end{minipage}\hspace{2pc}%
\begin{minipage}{3in}
\resizebox{3in}{!}{\includegraphics{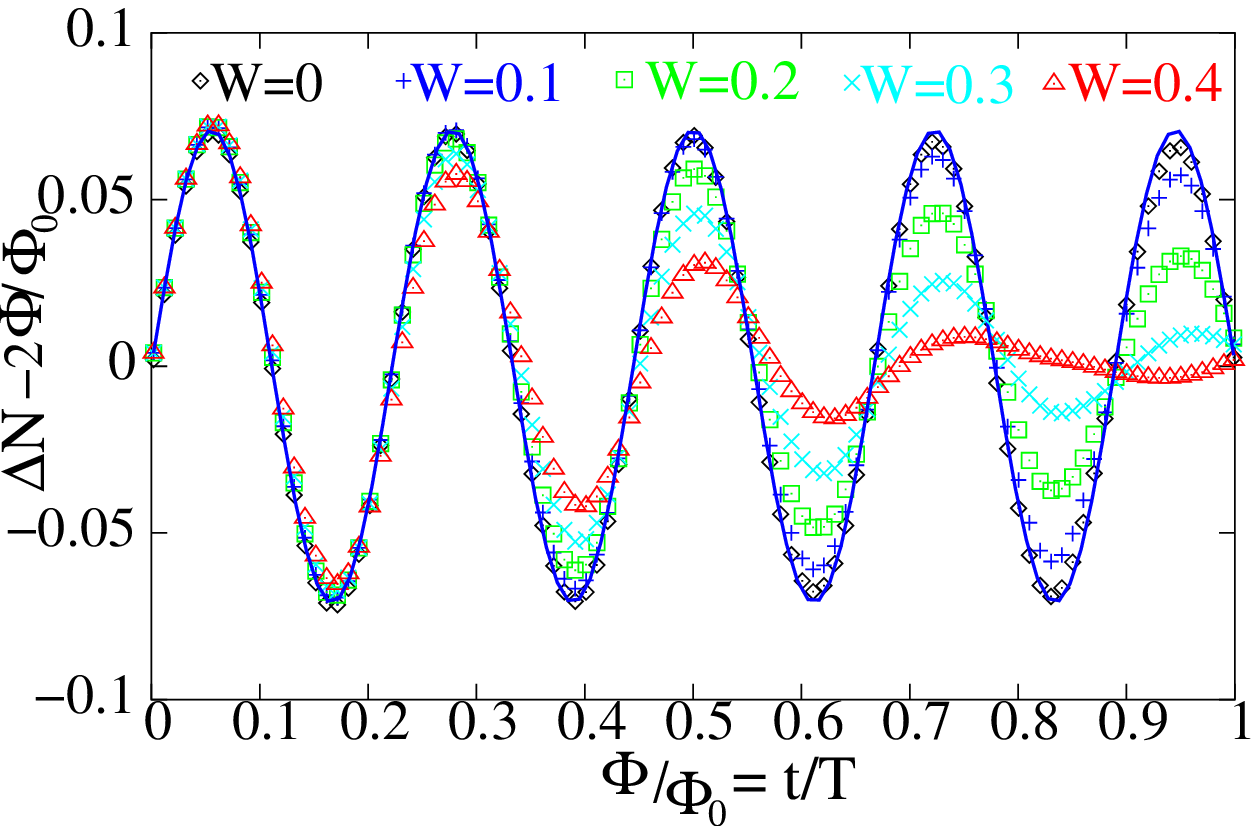}}
\caption{random-averaged $\Delta N- 2\Phi/\Phi_0$
as a function of $\Phi/\Phi_0$
for $H_{uni}$ on a cylindrical lattice
at $T=20$
with several strength of randomness $W$.
}
\label{fig:E}
\end{minipage}
\end{figure}
We found the result of $H_{imp}$ on the square lattice with the PBC
is quite similar to that of $H_{imp}$ on the cylindrical lattice.
It means that topology of two systems
can not affect edge-state pumping in the thermodynamic limit.
Moreover,
there is no singularity
at $\Phi/\Phi_0 = t/T=1$,
which is expected from Fig.~\ref{fig:Ac}.
Figure~\ref{fig:D}
shows that two regions, $t/T<1$ and $t/T>1$, are smoothly connected.
This result is similar to that on the cylindrical lattice again.

\paragraph{Effect of randomness}
Finally,
we studied the effect of random on-site potential $W$
as shown in Fig.~\ref{fig:E}.
Disagreement with the solid line (Eq.~\ref{eq:DeltaN})
becomes large with increasing $W$
and random-averaged $\Delta N$ approach $\Delta N=-2t/T$,
which is the form in the adiabatic limit.

\paragraph{conclusion}
In summary, we have studied numerically the non-adiabatic effect on 
edge-state pumping in the quantum Hall system.
The formula (Eq.~\ref{eq:DeltaN})
shows a good agreement with data of $H_{uni}$ on the cylindrical lattice
and connects quantized pumping $\Delta N(T)=-2$ in the adiabatic limit 
and no-pumping $\Delta N(T)=0$ in the sudden limit.
Non-adiabatic pumping can be efficient to increase net current but has the limitation
due to inhomogeneity of electric field.
We have observed clear steps of $\Delta N(t)$ due to the Landau gap
for $H_{uni}$ and for $H_{imp}$ at large $T>1/\omega_c$.
However, this effect is weak against randomness.

We acknowledge discussions with M. Kinuhara.
The computation in this work has been
 done using the facilities of the Supercomputer Center, Institute for
 Solid State Physics, University of Tokyo
and Altix3700BX2 at YITP in Kyoto
University.
This work has been supported
in part by Grants-in-Aid for Scientific Research,
No. 20340098, 20654034 from JSPS and 
No. 220029004, 20046002 on Priority Areas from MEXT.
\section*{References}
\bibliography{../macro,../wiki,../book}

\providecommand{\newblock}{}
\begin{thebibliography}{10}
\expandafter\ifx\csname url\endcsname\relax
  \def\url#1{{\tt #1}}\fi
\expandafter\ifx\csname urlprefix\endcsname\relax\def\urlprefix{URL }\fi
\providecommand{\eprint}[2][]{\url{#2}}

\bibitem{PRB.23.5632}
Laughlin R~B 1981 {\em Phys. Rev. B\/} {\bf 23} 5632

\bibitem{PRB.25.2185}
Halperin B~I 1982 {\em Phys. Rev. B\/} {\bf 25} 2185

\bibitem{PRL.49.405}
Thouless D~J, Kohmoto M, Nightingale M~P,  and den Nijs M 1982 {\em Phys. Rev.
  Lett.\/} {\bf 49} 405

\bibitem{PRL.71.3697}
Hatsugai Y 1993 {\em Phys. Rev. Lett.\/} {\bf 71} 3697

\bibitem{PRB.27.6083}
Thouless D~J 1983 {\em Phys. Rev. B\/} {\bf 27} 6083

\bibitem{SCIENCE.283.1864}
Altshuler B~L and Glazman L~I 1999  {\bf 283} 1864

\bibitem{SCIENCE.283.1905}
Switkes M, Marcus C~M, Campman K and Gossard A~C 1999  {\bf 283} 1905

\bibitem{PRB.77.153301}
Kaestner B, Kashcheyevs V, Amakawa S, Blumenthal M~D, Li L, Janssen T~J~B~M,
  Hein G, Pierz K, Weimann T, Siegner U and Schumacher H~W 2008 {\em Phys. Rev.
  B\/} {\bf 77} 153301

\bibitem{PRB.66.205320}
Moskalets M and Buttiker M 2002 {\em Phys. Rev. B\/} {\bf 66} 205320

\bibitem{PRL.95.130601}
Strass M, Hanggi P and Kohler S 2005 {\em Phys. Rev. Lett.\/} {\bf 95} 130601

\bibitem{PRB.33.6497}
Landauer R 1986 {\em Phys. Rev. B\/} {\bf 33} 6497

\bibitem{EPL.3.139}
de~Raedt H 1987 {\em Europhys. Lett.\/} {\bf 3} 139

\bibitem{PRB.51.10897}
Kawarabayashi T and Ohtsuki T 1995 {\em Phys. Rev. B\/} {\bf 51} 10897

\end{thebibliography}
\end{document}